# Runaway Electron Generation by Decelerating Streamers in Inhomogeneous Atmosphere

A Yu Starikovskiy[1,*] N L Aleksandrov[2], M.N.Shneider[1]

[1]Princeton University, Princeton, NJ08544, USA

[2]Moscow Institute of Physics and Technology, Dolgoprudny, 141700, Russia

[*]Author to whom any correspondence should be addressed

E-mail: astariko@princeton.edu

## Contents

# Abstract

The dynamics of positive and negative streamers is numerically simulated in atmospheric pressure air in the range of parameters corresponding to the streamer deceleration and termination in the middle of the discharge gap. A detailed comparison with experiments in air at constant and variable density demonstrates good agreement between the 2D simulation results and observations. It is shown that positive and negative streamers behave in radically different ways when decelerating and stopping. When the head potential drops, the negative streamer transits to the mode in which the propagation is due to the forward electron drift. In this case, the radius of the ionization wave front increases, whereas the electric field at the streamer head decreases further and the streamer stops. Its head diameter continues to increase due to the slow drift of free electrons in the residual under-breakdown field. On the contrary, the only advancement mechanism for a positive streamer with decreasing head potential is a decrease in the effective radius of the ionization wave, leading to a local increase in the electric field. This mechanism makes it possible to compensate for the decrease in the efficiency of gas photoionization at small head diameters. A qualitative 1D model is suggested to describe streamer deceleration and stopping for different discharge polarities. Estimates show that, during positive streamer stopping, the local electric field at the streamer head can exceed the threshold corresponding to the transition of electrons to the runaway mode when the head potential (relative to the surrounding space) decreases to ~1.2 kV in atmospheric pressure air. In this case, pulsed generation of a beam of runaway electrons directed into the channel of a stopping positive streamer can occur. The energy of the formed pulsed electron beam depends on the intensity of pohotoionization in front of the streamer head. This energy can vary from 700 V (when increasing the photoionization rate by a factor of 10 with respect to the value in atmospheric pressure air) to 2.6 kV (in air with an absorbing additive that reduces the photoionization rate by a factor of 1000). It is possible that this behavior of decelerating positive streamers can explain the observed bursts of X-ray radiation during the streamer propagation in long air gaps.

**Introduction**

There are remarkable natural phenomena in which streamer discharges develop in the inhomogeneous atmosphere. First of all, we can note red sprites, starters, and the streamer corona of blue jets and giant jets [1-4]. The density of the atmosphere falls exponentially with altitude, following the barometric formula, with a drop of *e* times approximately every ~ 7 km. The streamer structure observed in red sprites is similar to that observed in laboratory experiments, although the streamers in the red sprites are scaled by reduced air density in the upper atmosphere [1-2]. Since the lengths of sprites and jets are tens of kilometers, in the numerical modeling of their development and correct interpretation of the observation results, it is necessary to take into account the heterogeneity of the atmospheric density in height. Most of the numerical modeling and analytical analysis of streamer discharges have been carried out under the assumption of a constant gas density (see, for instance, reviews [5-7]). Rare exceptions are recent numerical simulations of streamer discharges in varying-density media, as applied to red sprites [2, 8-9] and streamer interaction with shock waves and other gas density nonuniformities [10-14].

To test computational models, it is important to compare the calculated results with observation data. When observing real sprites and jets, it is impossible to get detailed information about all parameters and conditions. Laboratory experiments provide such an opportunity. The results of the first attempt to simulate sprites in laboratory by using streamer discharges in a gradient density air was presented by Opaits et al. [15]. The purpose of the experiment was to obtain data that could be used for validation of numerical and analytical models. In this study, the controlled gradient density was created in a mixing air jet from a heat gun. Other methods of creating density gradients, such as supersonic nozzles and gas-filled tubes with controllable longitudinal thermal gradients, were also pointed out in [15]. In order to mimic red sprite conditions, Opaits et al. [15] studied streamer development from a region with reduced density to a region with higher gas density. It is also possible to conduct similar experiments when the discharge develops in the opposite direction, from a high-density region to a low-density region. This would allow simulation of the formation and development of blue jets, which propagate upward from the top of thunderclouds. Such an experiment has been recently made in laboratory [16].

It is worth noting that the parameters of streamers developing in an inhomogeneous atmosphere along the density gradient change in the course of propagation. It follows from similarity analysis [1-2] that, even for a constant potential of the streamer head, the head radius varies with the air density *n* approximately as $1/n$. This explains the observed narrowing of individual sprites during their propagation [1-2]. A similar dependence was shown in the laboratory experiment [15]. However, the interpretation of the results [15] is difficult since in these experiments the applied voltage was maintained constant, while the head potential decreased during the streamer propagation.

About 15 years ago, intense flashes of ultraviolet radiation excited in the earth's atmosphere were recorded by the Russian satellite Tatyana and the Taiwanese FORMOSA-2 [17-19]. Subsequently, flashes of ultraviolet radiation were repeatedly recorded by equipment installed on the near-Earth research Lomonosov satellite (an international designation MVL 300, or 2016-026A), launched in 2016 [20-21]. It was suggested and the corresponding estimates were made [22-23] that the flashes of ultraviolet radiation correspond to the 2nd positive system of nitrogen molecules, which are excited in the streamer corona of giant jets and sprites. The estimates showed good agreement with the data of satellite observations [17-19].

Flashes of hard gamma radiation of atmospheric origin [24-27], the so-called terrestrial gamma-ray flashes (TGF), were recorded from the board of artificial earth satellites along with ultraviolet flashes. The TGF burst duration ranged from several tens of microseconds to several milliseconds [28]. Among all natural high-energy phenomena occurring in the earth's atmosphere, TGFs generate the hardest photon spectrum, extending up to energies ~100 MeV [27].

Practically from the moment of the first registration of TGFs it was assumed that they are caused by bremsstrahlung of high-energy electrons of atmospheric origin [29]. Such relativistic electrons in the atmosphere, formed in the runaway mode, were predicted in [30]. Theoretical studies and numerical simulations have confirmed the possibility of a connection between the observed TGF flares and the deceleration of relativistic runaway electrons produced in thunderstorm electric fields (see, for example, [31]). At the same time, the origin of the initial beam of relativistic electrons without which the development of relativistic runaway electron avalanches would be impossible was not addressed.

Attempts have been made to explain the appearance of high-energy electrons during the propagation of negative streamers in the atmosphere without seed relativistic electrons, as well as during the collisions of streamers and during the development of the streamer zone ahead of leader tips [32-38]. The formation of high-energy electrons was assumed in relatively weak electrical fields ($E/n$ ~ 800-1000 Td) in the ionization wave of negative streamers. These fields are not sufficient for the generation of runaway electrons because typical energy losses are relatively small for high-energy electrons only. Thus, for $\varepsilon_e = 1$ MeV, the energy loss is primarily due to ionization and is equal to 2.09 keV/cm in atmospheric pressure air, which could be compensated by an electric field of $E_{comp} = 2.09$ kV/cm. For $\varepsilon_e = 100$ keV, one needs $E_{comp} = 4.55$ kV/cm. To compensate the energy losses for ionization at $\varepsilon_e = 10$ keV, the critical value of electric field is equal to $E_{comp} = 24.7$ kV/cm. Electrons with an initial energy of $\varepsilon_e = 1$ keV require $E_{comp} = 128$ kV/cm to maintain their energy at the same level [39]. "Cold" electrons during the acceleration process spend their energy on all types of excitations, including rotational, vibrational, and electronic excitation. That is why the electric field required to reach the run-away regime is much higher for low-energy electrons. Estimated critical value of electric field in this case is $E_{comp} > 450$ kV/cm in air at normal conditions ($E/n > 1700$ Td [40, 41]). This value is much higher than the typical electric fields in the streamer head and some special conditions must be satisfied to reach the runaway regime. These specific mechanisms of the extremely high electric fields and the runaway regime appearance were not considered in [32-38].

In this work a completely new mechanism was proposed to explaining the generation of runaway electrons in atmospheric discharges. Conditions were found when the electric field at the streamer head exceeds the breakdown threshold by a factor of 30-50 and significantly exceeds the critical values for electron runaway. As a result, the majority of electrons could transform into the runaway regime. It was shown that positive streamer deceleration in the undisturbed atmosphere by itself can generate extremely high electric fields exceeding the runaway threshold. In such fields, intensive relativistic electron beams, as well as electron-induced X-ray radiation, could be formed. The effect under study depended on the discharge polarity, which could be verified experimentally or by analyzing available observation data.

**Numerical modeling**

Streamer deceleration in the atmosphere was simulated for both discharge polarities. The initiation and propagation of streamers were investigated using an axisymmetric two-dimensional fluid model. The model and computational method were presented in detail in previous papers [42-46].

The transport and reactions of charged particles were described in the local approximation, while the non-local approach was utilized to describe photoionization generating seed electrons in front of the streamer head by means of ionizing radiation emitted from the head and streamer channel:

$$\frac{\partial n_e}{\partial t} + \text{div}(\vec{v_e} \cdot n_e) = S_{ion} + S_{photo} - S_{att} - S_{rec}^{ei} \tag{1}$$

$$\frac{\partial n_p}{\partial t} = S_{ion} + S_{photo} - S_{rec}^{ei} - S_{rec}^{ii} \tag{2}$$

$$\frac{\partial n_n}{\partial t} = S_{att} - S_{rec}^{ii} \tag{3}$$

$$\Delta \varphi = -\frac{e}{\varepsilon_0}(n_p - n_e - n_n) \tag{4}$$

where $S_{ion}$ is the ionization rate, $S_{photo}$ is the rate of photoionization, $S_{att}$ is the rate of electron attachment, and $S_{rec}^{ei}$ and $S_{rec}^{ii}$ are the rates of electron-ion and ion-ion recombination, respectively. The charged species under consideration were electrons, $O_2^+$, $N_2^+$, $O_4^+$, $O_2^+N_2$, $N_4^+$, and $O_2^-$. The electron transport and rate coefficients were calculated from a numerical solution of the Boltzmann equation in the two-term approximation. The kinetic model and rate coefficients used were described in detail elsewhere [44, 46]. The local electric field approximation for electrons is the classical approach in fluid models. More sophisticated methods such as extended (high-order) fluid models and hybrid models have been suggested to simulate streamer properties [7, 47-51]. These methods are more adequate for describing energetic and run away electrons. However, they are time-consuming and difficult to use for simulating long streamers. Therefore, to simulate positive and negative streamers in a long (10 cm) air gap, we utilized the classical fluid model with the local electric field approximation allowing reasonable run time and memory consumption.

The system of equations (1) - (3) was converted into a finite-difference form by the control volume method. Equations (1) and (2) were numerically solved by the splitting of physical processes. The transport of charged particles was calculated using an explicit finite-difference backward approximation of the first order accuracy in space and time, whereas the right-hand sides of the equations were calculated using the Euler technique of the first order in time [42]. To solve Poisson's equation (4), the Gauss–Seidel overrelaxation technique was utilized.

The photoionization source term in equations (1)-(2) was calculated using the approach [52]:

$$S_{photo} = \frac{1}{4\pi} \frac{p_q}{p+p_q} \int_V d^3 r_1 \frac{S_{ion}(r_1)}{|r-r_1|^2} \Psi(|r-r_1|p), \qquad (5)$$

where $p$ is the gas pressure, $p_q$ is the quenching pressure ($p_q = 30$ Torr for air [52]), $\Psi(|r-r_1|p)$ is the coefficient of absorption for the ionizing radiation in the gas. Integral equation (5) was solved directly both for positive and negative streamers. To reduce the computation time, we calculated the VUV photon flux using a coarse mesh and averaging photon generation and absorption over 10 point regions.

The computational mesh was adapted as the peak of the electric field moved in the discharge gap. The typical number of cells was $N_z \times N_r = 4096 \times 256$. The minimum spatial step in the radial and axial directions was $\delta r_{min} = \delta z_{min} = 1.5$ μm. In all cases, an adaptive computational grid was used with a uniform step along the discharge axis and radius near the leading ionization wave and the high-voltage electrode tip. The grid spacing increased exponentially with increasing distance from the region of the maximum ionization rate. The time step was, as a rule, around $\delta t \cong 5 \times 10^{-14}$ s, which was required for the stability and accuracy of the numerical scheme.

The calculations were carried out for two electrode geometries at different voltages across the discharge gap in order to exclude the influence of computational effects on the results and conclusions of the work. As noted in [53], the development of streamers of any polarity is determined by the same set of processes. They are electron impact ionization, photoionization, and electron drift in a self-consistent electric field. The difference between the positive and negative discharges is small when the applied voltage is much higher than the threshold values for ionization wave development. It will be shown below that this difference becomes considerable when the applied voltage is close to the threshold values.

**Point-to-plane geometry, low voltage discharge: experiment and modeling**

First we considered streamer development under the conditions corresponding to the experimental study of the propagation length and effective radius of the streamer discharge channel in the needle-to-plane geometry [15]. The experiment was made in a uniform air flow at atmospheric pressure and in an air flow with a density gradient (Figure 1). The peak voltage at the high-voltage electrode was 5.4 kV, the voltage rise time was 50 ns, and the pulse length exceeded 500 μs. The distance between the electrodes was 35 mm.

From Figure 1, the propagation length of a positive streamer in atmospheric-pressure uniform air is short, around 8-10 mm (case A). In non-uniform air, a twofold decrease in the gas density near the high-voltage electrode (case B) allows the streamer to almost double the path length, in spite of the fact that the gas density near the flat low-voltage electrode becomes close to normal. The diameter of the streamer channel decreases with the distance from the high-voltage electrode.

The experimental data [15] presented in Figure 1 make it possible to validate the numerical model used in this work to simulate a streamer discharge in the deceleration regime.

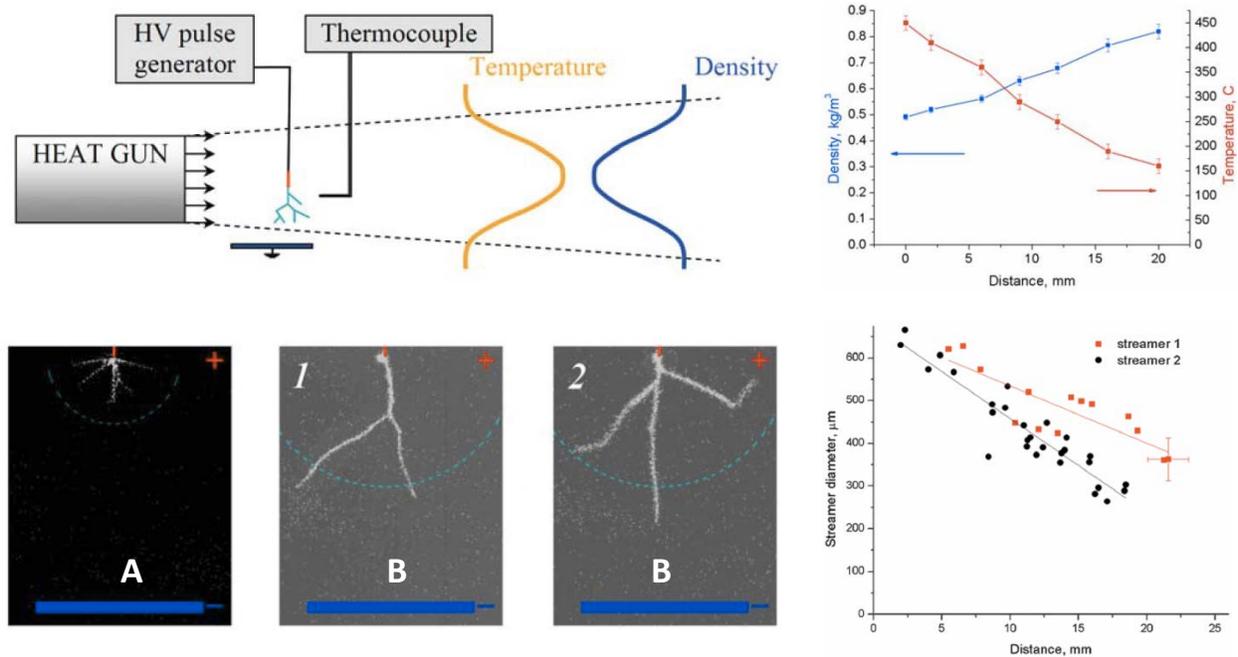

Figure 1. Discharge gap geometry, gas density distribution, and streamer propagation length under experimental conditions [15]. Upper row. Left: experimental setup; Right: temperature and gas density distribution along the discharge axis. Lower row. Left: Discharge development (A) in uniform density room air and (B) in gradient density air. The lowest density is on the top near the pin electrode, whereas the highest density is on the bottom near the plate electrode. Image size 32×40 mm. Right: Diameter of streamer channel versus distance in gradient density air for cases (1) and (2).

The parameters for comparison are the limiting channel length for the discharge development in the homogeneous and inhomogeneous gases, as well as the discharge radiation profile, which determines the observed effective channel radius. On a short-time scale, the second positive system of nitrogen is the most effective source of the streamer discharge radiation in air [53]. However, in [15], integral imaging was used in which the contribution of the first positive system to the overall emission was significantly increased. In addition, due to the limited spatial resolution of the ICCD cameras (as a rule, the entire system resolution did not exceed 500×500 pixels per frame), even in the integral imaging mode, it is fundamentally impossible to measure the geometric parameters of an object with a length/radius ratio of more than several tens. For example, with an image size of 40 mm, the resolution of an ICCD camera in any direction, in principle, cannot be better than 0.1 mm. Therefore, measurements [15] of the streamer channel radius inevitably overestimate the real values.

**Positive streamer deceleration**

The simulation of positive streamer deceleration was carried out under conditions that correspond to the experiment [15]. The high-voltage pin electrode was located in the low-density gas region. The plain ground electrode was placed in the normal-density air at a distance of 30 mm from the pin electrode. The gas density profile in the discharge gap corresponded to the measured spatial density distribution (Figure 1, B). The radius of curvature of the pin electrode was set to 40 microns. The rise time of the voltage pulse was 50 ns from 0.1 to 5.4 kV where the voltage increased linearly on the high-voltage electrode.

The results of calculations are presented in Figures 2 and 3. Figure 2 shows the result of the calculation for the case of uniform air (case A in Figure 1). The left part of the figure shows the temporal evolution of the axial profiles of the potential, electric field strength, and electron concentration in the discharge gap. The time is reckoned from the beginning of the voltage pulse. The right column shows the spatial distributions of the electric field and electron concentration at different moments.

The potential of the streamer channel near the high-voltage electrode rises during the first 50 nanoseconds. The streamer starts at almost maximum voltage. This result is in good agreement with the observations [15] where the minimum possible voltages were specially used to study streamer termination. Then, an ionization wave starts from the high-voltage electrode and carries the potential into the discharging gap. From Figure 2, the potential of the streamer head decreases rapidly due to the finite resistance of the streamer channel. In a uniform flow, the average electric field in the channel is $U/L \sim 3.25$ kV/cm. The decrease in the head potential leads to a decrease in the streamer velocity. The streamer reaches its limiting length in ~90 ns. The calculated limiting length of the streamer channel in this case is $L_{max} \sim 9$ mm, also in agreement with the experimental data [15].

The radius of the channel changes as the streamer propagates through the discharge gap. As the head potential decreases, the channel radius drops sharply (Figure 2, right column). Such a decrease in the channel radius is completely natural, since in order to maintain a high ionization rate in the leading ionization wave, the streamer must maintain a high value of the electric field at the head. Simultaneously with a decrease in the head potential due to the voltage losses in the channel, the progress of the streamer in the discharge gap should inevitably lead to a decrease in the radius of curvature of the leading ionization wave.

However, it is follows from Figure 2 that the electric field at the streamer head does not remain constant during the deceleration. The field increases from 20 MV/m ($E/n \sim 745$ Td) at the beginning of the streamer movement to almost 50 MV/m ($E/n \sim 1860$ Td) when approaching its stopping point. A significant increase in the electric field at the head is a necessary condition for the continuation of the positive streamer motion due to a decrease in the photoionization efficiency. As the streamer propagates in the gap, a decrease in the radius of curvature of the streamer head leads to the fact that the path length of the pre-ionizing VUV radiation becomes greater than the radius of the head. As a result, seed photo-electrons necessary for the development of the ionization wave [53] are generated in the regions with relatively weak electric fields. This leads to the formation of weak avalanches and hence slows down the radial expansion of the channel. At the same time, the electron concentration on the channel axis sharply increases, since the absolute value of the electric field at the head increases. The radius of the conductivity of the channel, however, drops sharply in this case; therefore, the conductivity of the entire channel does not significantly increase. The field in the channel immediately behind the head is almost halved as the streamer approaches the stopping point (Figure 2). Thus, a decrease in the potential of the head of a positive streamer during its propagation in the discharge gap leads to the development of instability near the stopping point. This instability demonstrates a decrease in the head radius and an increase in the electric field at the head to values comparable to the electron runaway threshold.

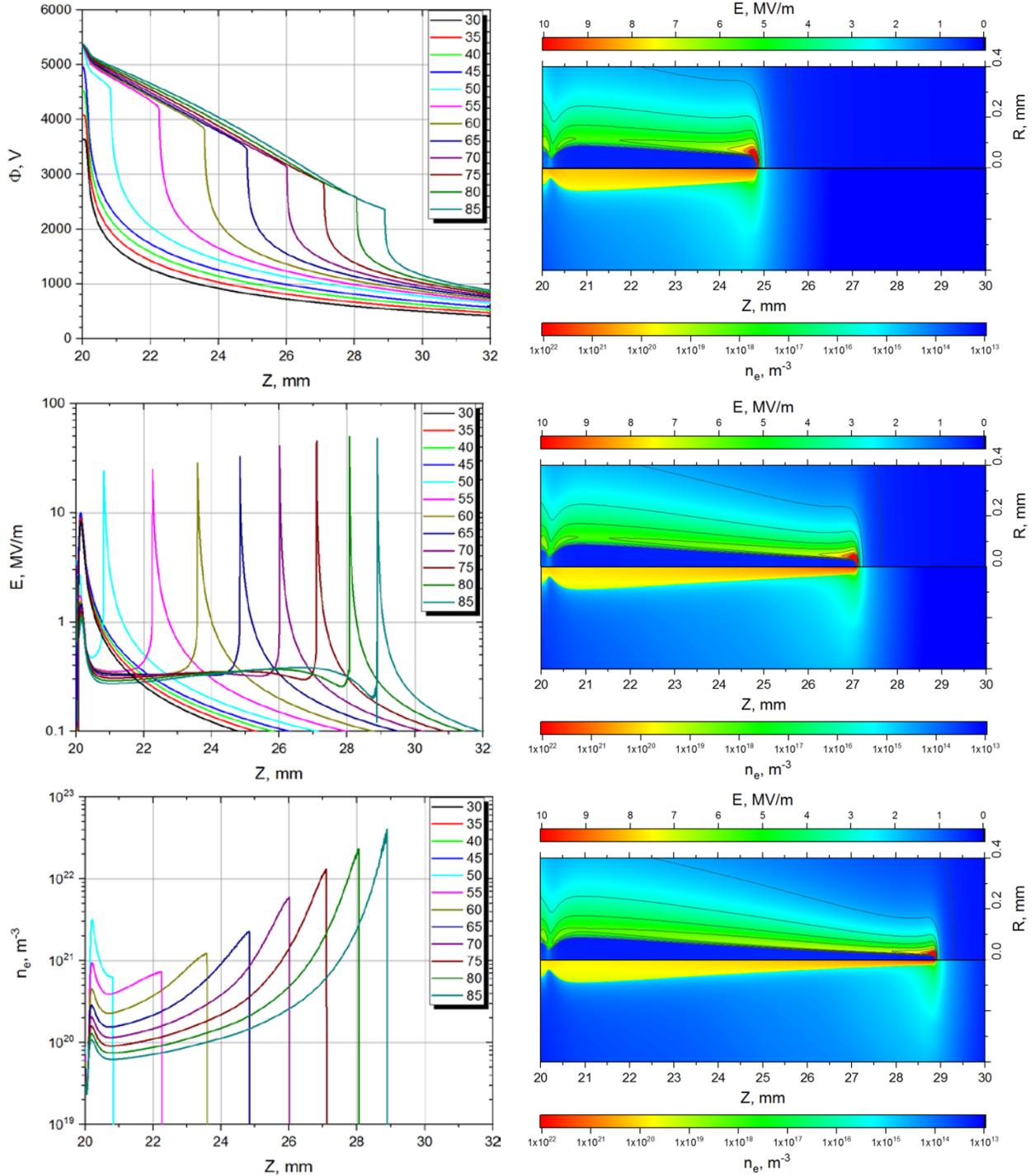

Figure 2. Development of a positive streamer in uniform air. $U_{max}$ = +5.4 kV, $P$ = 760 Torr, $T$ = 300 K. Left column: axial profiles of potential, electric field, and electron concentration on the channel axis at different times. The time scale is shown in nanoseconds. Right column: spatial distributions of electric field and electron concentration at $t$ = 65, 75 and 85 ns after the high-voltage pulse start.

Note that our simulation of streamer development at $E/n > 2$ kTd is unjustified because the local drift-diffusion model used in this work is not acceptable here. However, in the case of a positive streamer, there is no physical mechanism that could significantly slow down the development of the instability of the leading ionization wave when the head radius decreases to values less than the path length of the preionizing VUV radiation. The possible formation of runaway electrons in a strong field at the front of the leading ionization wave has practically no effect on the streamer development, since such electrons move towards the streamer channel and do not affect the conditions in front of the streamer head.

In this case, deceleration of the positive streamer can lead to the local formation of a region of ultrahigh electric fields and the generation of relativistic electrons with an energy comparable to the potential of the streamer head at the moment of its stopping.

In the next sections, we will analyze numerically and analytically the possibility of developing such a scenario in various conditions.

Figure 3 shows the propagation of a streamer in the gap at a variable gas density (corresponds to experiments "B" in Figure 1). In this regime, the gas density near the tip of the high-voltage electrode is half the normal one. The discharge starts at almost half the peak voltage 20-25 ns after the beginning of the voltage pulse. An increase in the potential of the high-voltage electrode to the 50$^{th}$ nanosecond raises the streamer velocity. At the same time, the potential of the head also increases (Figure 3, top line, left column). In this case, the streamer channel has an almost constant diameter along its length (Figure 3, top line, right column). Further development of the discharge occurs at a constant potential of the high-voltage electrode and an inevitable decrease in the potential of the streamer head due to voltage losses in the channel. In the case of streamer motion through a rarefied gas, the voltage drop in the channel is significantly smaller than that in the case of a gas of constant density. As a result, the average electric field in the streamer channel developing from the low-density region is $U/L \sim 1.45$ kV/cm, approximately half the value of $U/L$ in uniform air. The streamer decelerates sharply as it moves through the gap. This is a consequence of both a decrease in the head potential and an increase in the gas density as the streamer approaches the grounded electrode (Figure 1). As in the case of uniform air, a decrease in the head potential leads to a decrease in its radius, an increase in the electric field at the leading ionization wave, and an increase in the electron concentration on the axis of the streamer channel (Figure 3).

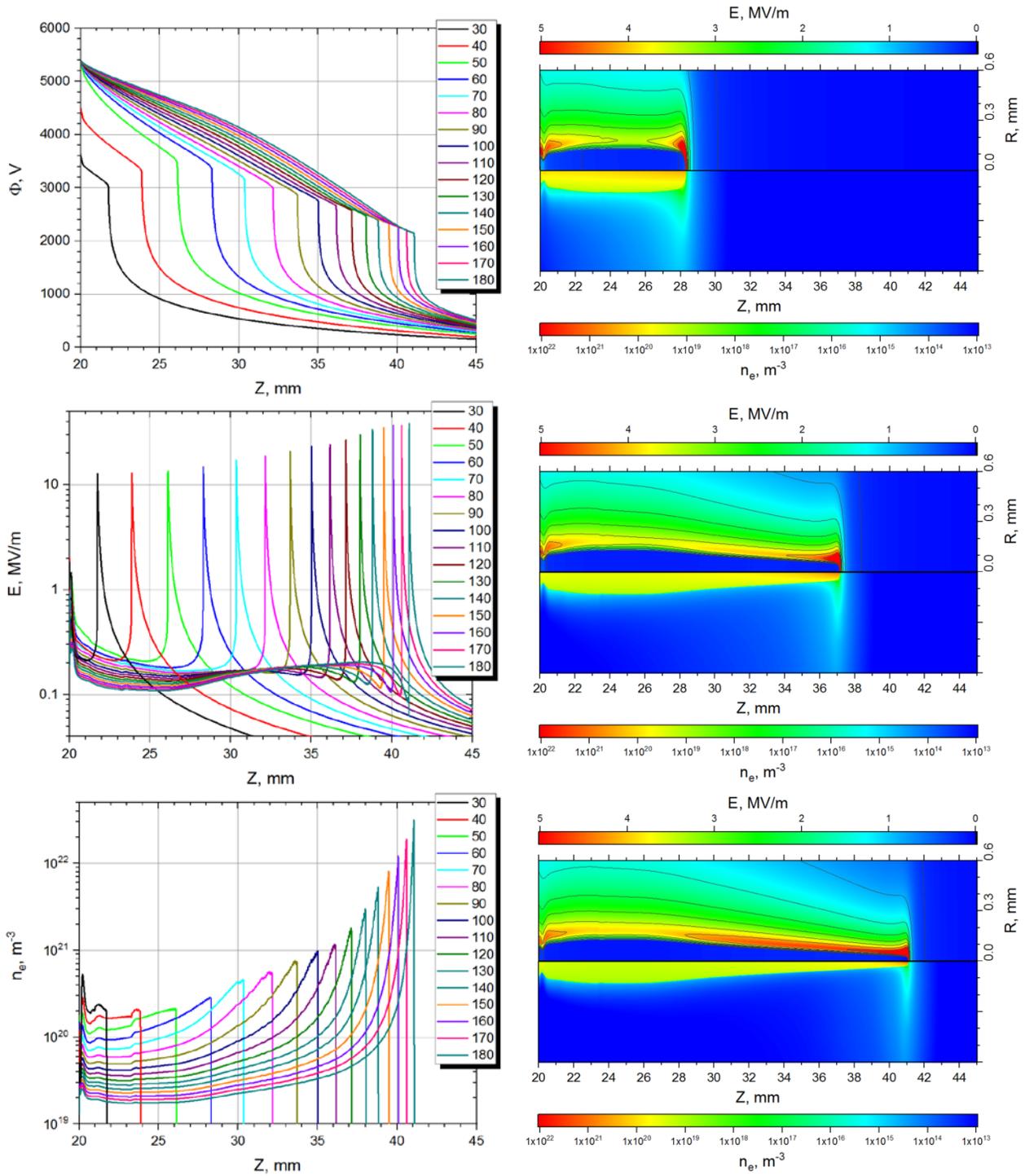

Figure 3. Development of a positive streamer in non-uniform air. $U_{max}$ = +5.4 kV, $P$ = 760 Torr, $T$ = 300-720 K. The density profile corresponds to the data in Figure 1. Left column: axial profiles of potential, electric field and concentration of electrons on the channel axis at different times. The time scale is shown in nanoseconds. Right column: spatial distributions of electric field and electron concentration at $t$ = 60, 120 and 180 ns after the high-voltage pulse start.

From Figure 1, the streamer limiting length was from 18 to 22 mm [15]. Almost the same limiting length is obtained in the calculation (Figure 3). Good agreement between the calculated and measured values of this key streamer parameter indicates a fairly accurate account of the key mechanisms and processes in the numerical model.

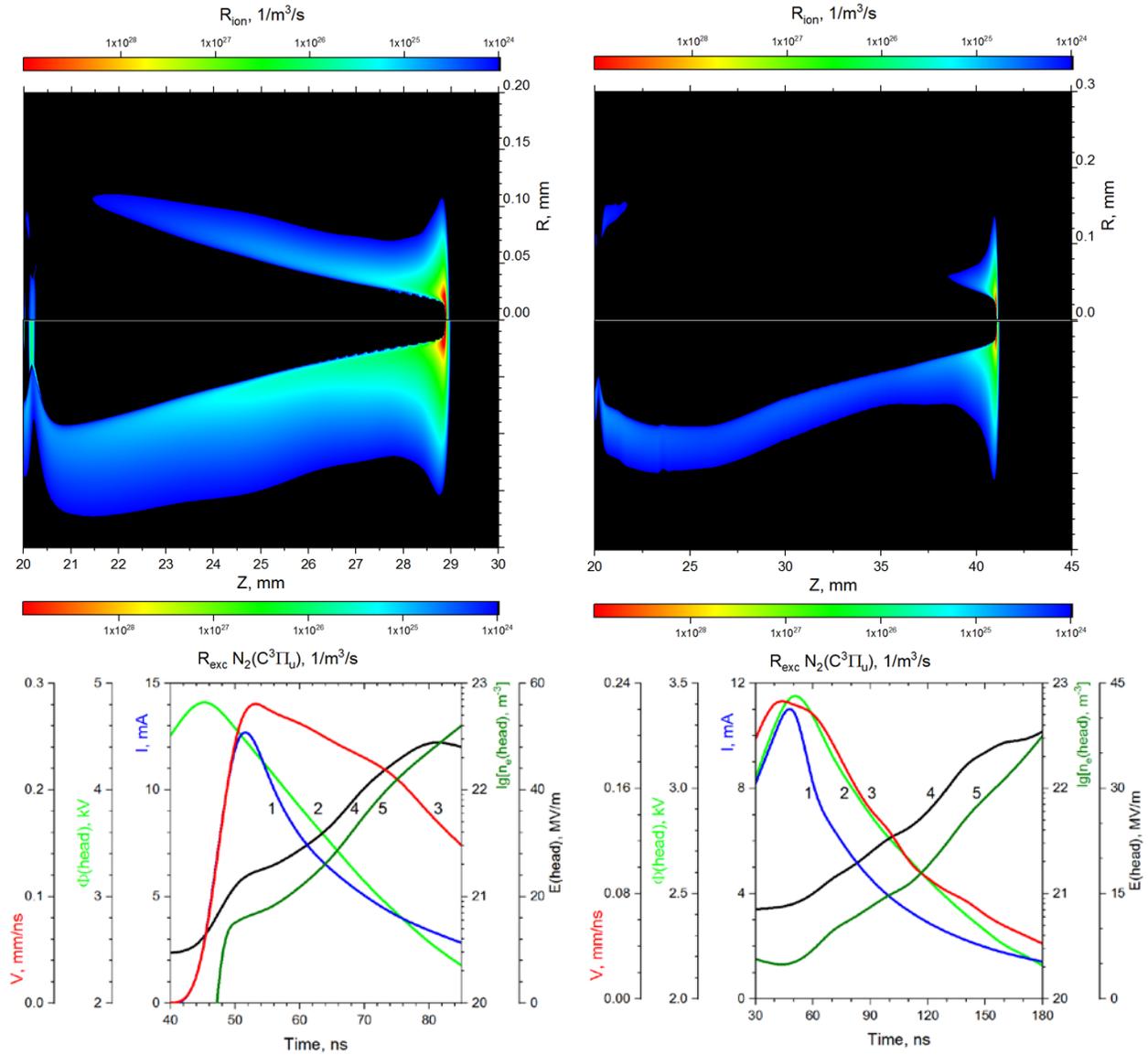

Figure 4. Deceleration of a positive streamer in air. Left: development in uniform air; right: streamer propagates from a low-density region to a high-density region. Upper row: spatial distributions of ionization rate (upper half of the figure) and electron-impact excitation rate for the upper level of the second positive nitrogen system (lower half of the figure). The distributions are plotted for the time instants close to the complete streamer stop (80 and 180 ns, respectively). Bottom row: temporal evolution of propagation velocity, streamer current, head potential, head electric field, and peak electron concentration at the discharge axis.

Figure 4 compares the results of calculating the deceleration of a positive streamer in uniform and non-uniform air in the point-to-plane geometry. The left column demonstrates the development of a discharge at constant gas density, whereas the right column shows the propagation of a streamer from the rarefied gas region to the dense gas region (the conditions correspond to Figure 1). The upper line of the figure shows the distribution of the gas ionization rate (upper half of the figure) and the electron-impact excitation rate for the upper level of the second positive system of nitrogen (lower half of the figure).

The distributions are plotted for a time instant close to the complete stop of the streamer (80 and 180 ns, respectively). It is clearly seen that the cone-shaped profile of the streamer channel recorded in the experiment is fully reproduced in the calculation. Calculated data show that the observed diameter of the streamer channel narrows for short times (a) from 200 to 40 microns in the case of discharge deceleration in uniform air and (b) from 320 to 60 microns in the case of the development of a discharge from a rarefied gas to a dense one. These data are in agreement with the experimental results [15], taking into account the limitations on the spatial resolution across the streamer channel for the diagnostic system used in the experiment.

The bottom row of diagrams in Figure 4 shows the temporal evolution of parameters of the streamer head as it moves through the gap. The potential of the streamer head (curve (2)) decreases rapidly due to the increasing voltage losses in the streamer channel. Because of this, the streamer velocity (curve (3)) and current (curve (1)) decrease with time. However, in both uniform and non-uniform cases, the peak electric field at the streamer head (curve (4)) and the maximum electron concentration at the channel axis (curve (5)) sharply increase with time. The magnitude of the local reduced electric field at the streamer head reaches $E/N \sim$ 1900 Td for both cases under consideration. Further simulation in the approximations used would not change the tendencies towards an increase in the magnitudes of the local electric field and the electron concentration in the streamer head. However, this calculation would become physically incorrect due to the inapplicability of the local drift-diffusion approach under conditions of strong electric fields and sharp gradients of all major parameters [54-55].

### Negative streamer deceleration

To compare the development of discharges of different polarity, the streamer deceleration in the same geometry was also simulated for negative voltage pulses. It is well known that a negative streamer cannot develop under the same voltage as a positive one [5]. Preliminary calculations showed that the negative streamer reaches the same length as the positive one at an initial voltage $U = -20$ kV. To reduce the computation time, the voltage rise time was reduced to 1 ns.

Figures 5 and 6 show the calculated results for the development of a negative streamer in the point-to-plane geometry. Figure 5 shows the development of a discharge in uniform air, whereas Figure 6 demonstrates its development from the low-density region to the high-density region (similar to Figure 1). At low initial voltages, the behavior of the negative streamer differs greatly from that of the positive streamer. The head potential the negative streamer begins to decrease immediately after its start (Figure 5). However, as opposed to the case of positive polarity, the field at the streamer head and the electron concentration rapidly decrease rather than increase. This is associated with a rapid increase in the radius of the negative streamer channel due to electron drift in the direction of the ionization wave development. Figure 5 (right column) shows the spatial distributions of the electric field and electron concentration in the discharge gap at different times. Already 10 nanoseconds after the start, the head radius of the negative streamer reaches 2 mm, and the maximum electric field drops to 4 MV/m. Under similar conditions, the electric field at the head of the positive streamer was 15 MV/m, despite the significantly lower potential of the head.

The obtained difference in the dynamics of the positive and negative discharges is associated with the opposite directions of electron drift near the streamer head. In [53], the effect of voltage polarity was studied for "strong" streamers propagating at high velocities, much higher than the electron drift velocity at the maximum field ($V_{str} \gg V_{dr}$). The main mechanism of development of both positive and negative "strong" discharges was gas preionization in front of the leading ionization wave by VUV radiation generated at the wave front.

As opposed to simulating "strong" streamers [53], this paper addresses the properties of "weak" streamers, whose propagation velocity is comparable to the electron drift velocity at the ionization wave front. This mode is realized when the streamer decelerates and comes to a halt. Here, the leading mechanism of streamer propagation changes for negative polarity. Comparable velocities of electron drift and streamer propagation mean that electrons born ahead of the ionization wave front are practically fixed relative to this front for a long period of time. Therefore, a high degree of electron multiplication is obtained even in not-too-high electric fields and the negative streamer channel widens due to the radial expansion caused by the electron drift.

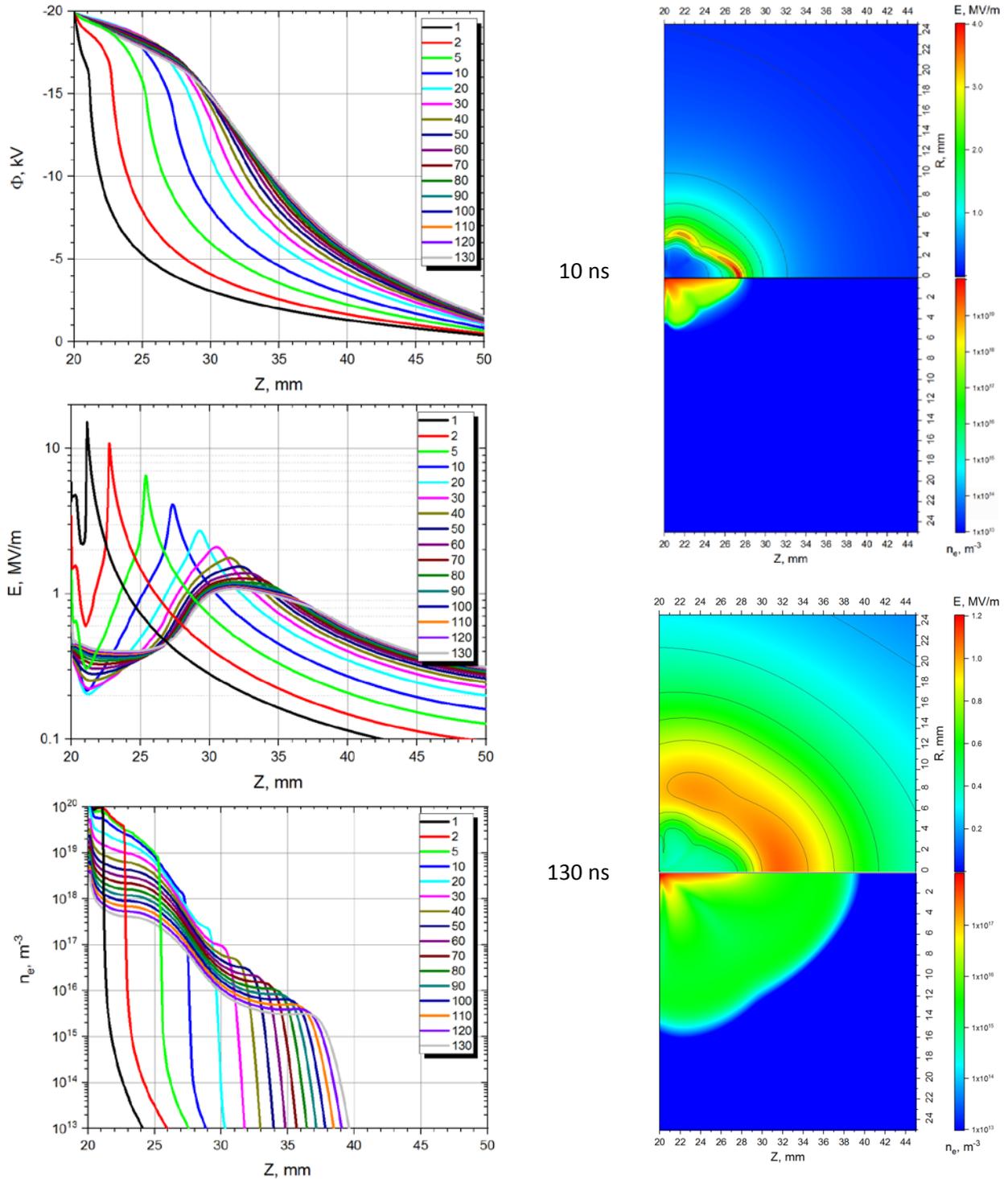

Figure 5. Development of a negative streamer in uniform air. $U_{max}$ = -20 kV, $P$ = 760 Torr, $T$ = 300 K. Left column: axial profiles of potential, electric field, and electron concentration on the channel axis at different times. The time scale is shown in nanoseconds. Right column: spatial distributions of electric field and electron concentration at $t$ = 10 and 130 ns after the high-voltage pulse start.

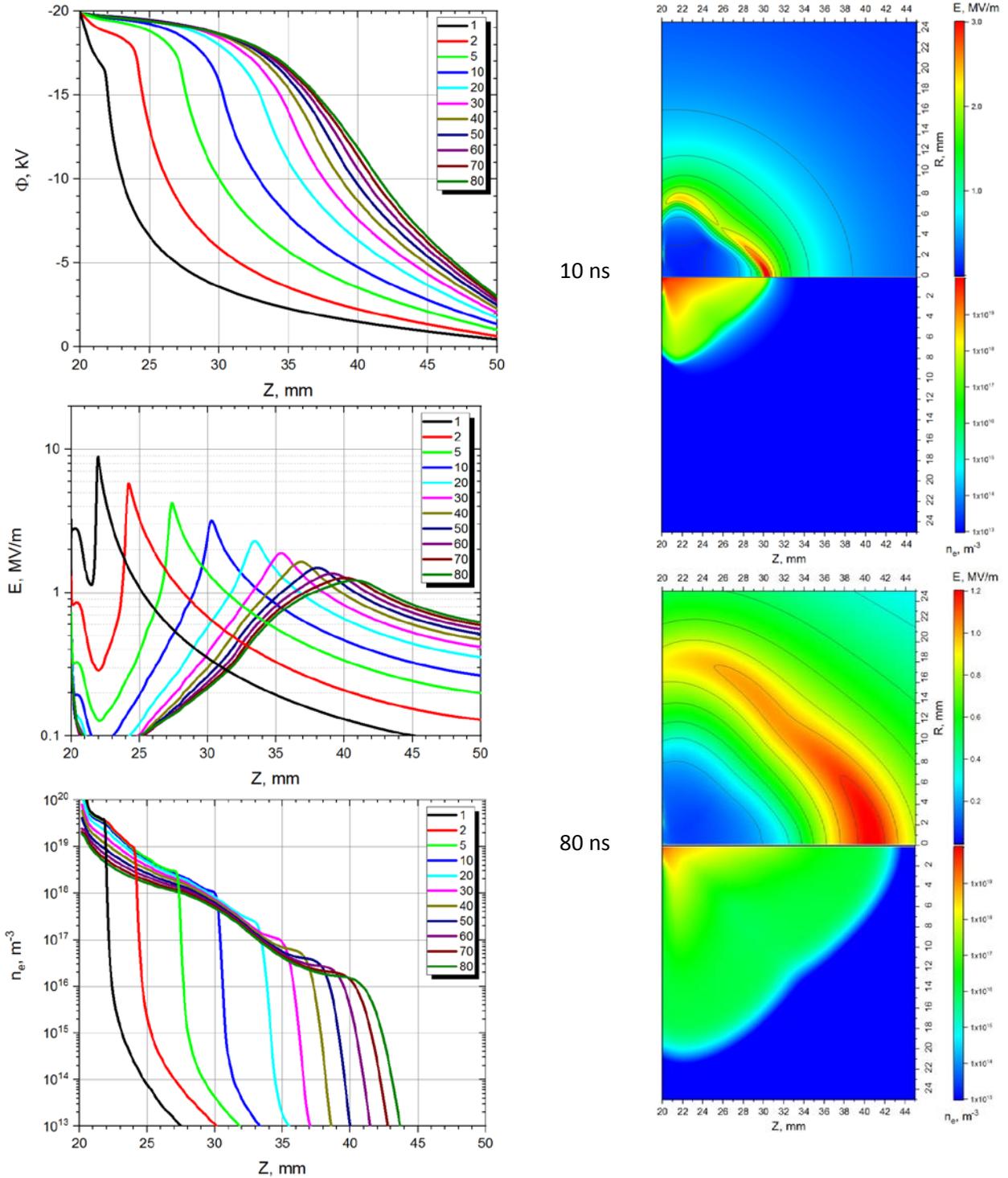

Figure: 6. Development of a negative streamer in non-uniform air. $U_{max}$ = -20 kV, $P$ = 760 Torr, $T$ = 300-720 K. The density profile corresponds to the data in Figure 1. Left column: axial profiles of potential, electric field and electron concentration on the channel axis at different times. The time scale is shown in nanoseconds. Right column: spatial distributions of electric field and electron concentration at $t$ = 10 and 80 ns after high-voltage pulse start.

The calculated results for the development of a negative streamer in non0uniform air (similar to positive discharge in Figure 3) show a similar behavior (Figure 6). The ionization wave starts from the high-voltage electrode relatively quickly, but at the same time the wave develops with a high radial expansion rate. As a result, despite a much more moderate decrease in the potential of the streamer during its propagation along the gap, a rapid decrease in the peak electric field at the head is observed (Figure 6, left column). The electron concentration in the streamer head rapidly drops as it develops through the gap, and the streamer slows down sharply. From Figure 6 (right column), the spatial distributions of the electric field and electron concentration at $t = 80$ ns show that the ionization wave has reached a radius of almost 15 mm by this instant, and its velocity has dropped from 2 to 0.1 mm/ns. At this moment, the peak electric field at the wave front has already dropped to $E = 1.2$ MV/m, which corresponds at this point to the subcritical reduced electric field $E/n \sim 56$ Td. Further progress of the streamer stops, and slow (on a given time scale) electron recombination and attachment lead to a gradual decay of conductivity in the discharge region. The spatial charge injected by the streamer in the discharge gap is carried by negative ions with low mobility. This charge suppresses the development of a secondary streamer.

**Plane-to-plane geometry, high voltage discharge: modeling**

For a more detailed analysis of various modes of the streamer deceleration, calculations were also carried out in the plane-to-plane geometry, with an increased distance between the electrodes and higher applied voltage. The geometry was similar to that used previously for simulating "strong" positive and negative streamers [53]. We considered a single streamer developing in a 14 cm plane-to-plane gap. All calculations were made for air at atmospheric pressure and room gas temperature. The computational domain was $150 \times 150$ mm$^2$. The high-voltage electrode was a plate at $Z = 10$ mm with a semi-ellipsoidal needle (large semi-axis 8 mm and small semi-axis 0.8 mm) protruding from the plate center. The streamer was initiated near the tip of the needle and propagated along the axis of the gap to the grounded electrode.

As opposed to the conditions studied previously [53], relatively low voltages were chosen, which ensured streamer termination approximately in the center of the discharge gap. As in the case of the point-to-plane geometry, the critical voltage of positive streamer initiation in the plane-to-plane geometry is much less than that of negative streamer initiation. For the same other conditions, the streamer limiting length was about 7 cm with a positive pulse amplitude of $U = +60$ kV and a negative pulse amplitude of $U = -150$ kV.

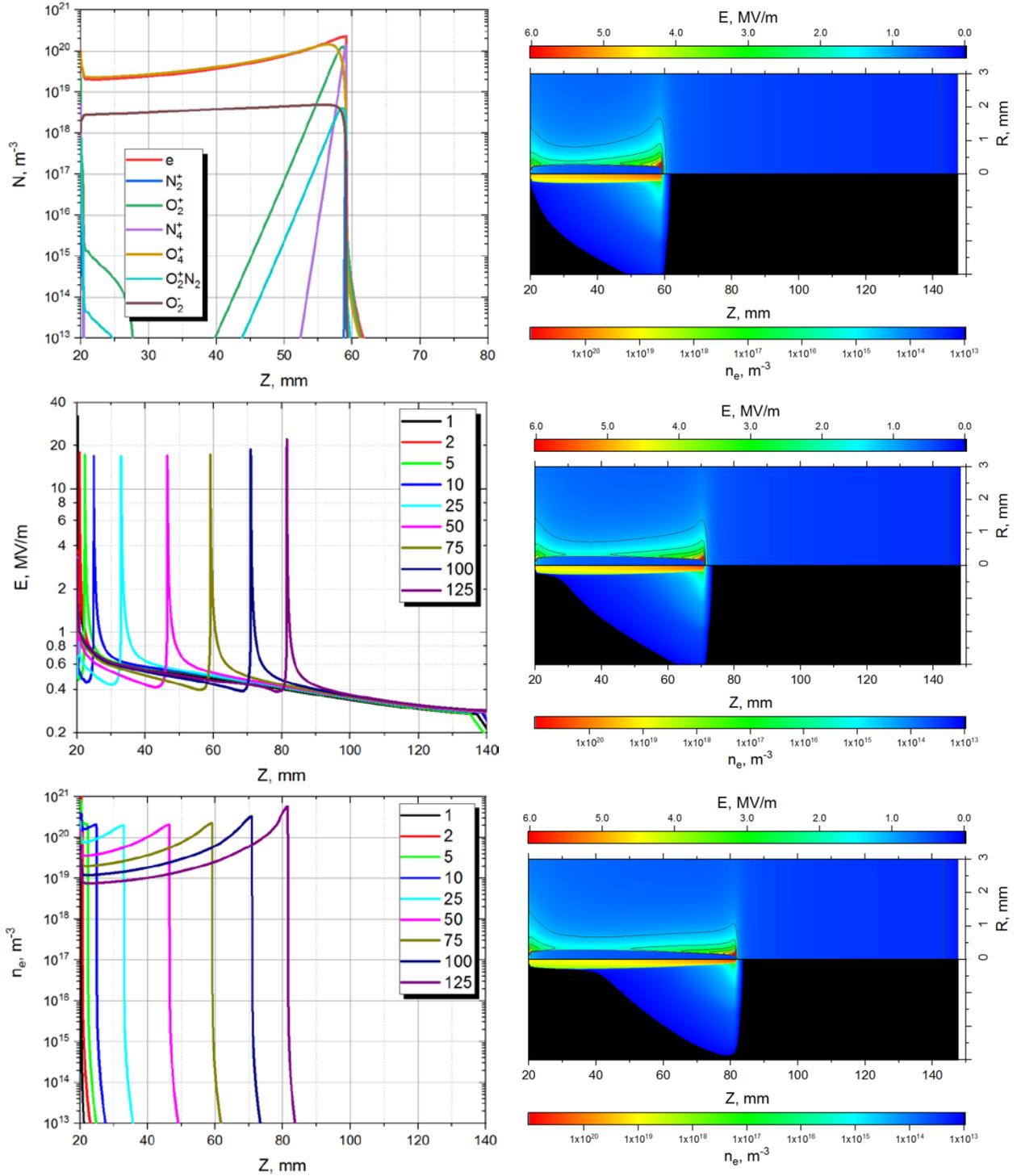

Figure 7. Development of a positive streamer in uniform air. $U_{max} = +60$ kV, $P = 760$ Torr, $T = 300$ K. Left column: axial profiles of the concentrations of the main ions at $t = 75$ ns and temporal evolution of the axial profiles of electric field and electron concentration. The time scale is shown in nanoseconds. Right column: spatial distributions of electric field and electron concentration at $t = 75$, 100 and 125 ns.

Figure 7 shows the temporal evolution of the axial profiles of potential, electric field, and electron concentration on the axis of a positive streamer. The time of streamer development is relatively long because of the high values of the gap length and high applied voltage. In this case, electron-ion recombination has time to become important for the plasma decay in the streamer channel (Fig. 7, top left column). The concentration of free electrons decreases by almost an order of magnitude for 75 ns. Positive molecular ions are converted into cluster ions in the sequence $N_2^+ \rightarrow N_4^+ \rightarrow O_2^+ \rightarrow O_4^+$, accelerating the electron-ion dissociative recombination and plasma decay. The electron attachment to molecular oxygen is slow, and the concentration of negative oxygen ions remains relatively low (Figure 7). However, despite the significant differences in the configuration of the discharge gap, the discharge voltage, and the times of streamer deceleration, the main results remain the same as those obtained for the development of a low-voltage discharge in the point-to-plane geometry (Figure 2).

A pronounced decrease in the potential of the streamer head leads to a decrease in the radius of curvature of the ionization wave and an increase in the peak values of the electric field and electron concentration at the streamer head. The streamer velocity decreases, since the total efficiency of the ionization and photoionization processes decreases when the size of the streamer head becomes comparable to or less than the path length of the ionizing VUV radiation.

The development of a negative streamer in the plane-to-plane geometry (Figure 8) also follows a scenario similar to the dynamics of the development of a negative streamer in the point-to-plane discharge gap (Figure 5). An increase in the radius of curvature of the leading ionization wave, a sharp decrease in the peak electric field at the front, and a decrease in the electron concentration in the streamer head are also observed. As in the case of a high-voltage positive streamer, fast dissociative recombination between electrons and cluster ions also occurs in the negative discharge channel. Due to the long duration of the discharge development, it takes longer time for the transition from the electron-ion plasma to the ion-ion one. This almost completely stops the charge redistribution in the gap. The space charge remaining in the region of the stopped head of the negative streamer is almost completely carried by $O_2^-$ ions. The peak electric field is only $E = 1.5$ MV/m, which corresponds to a deeply subcritical reduced field of

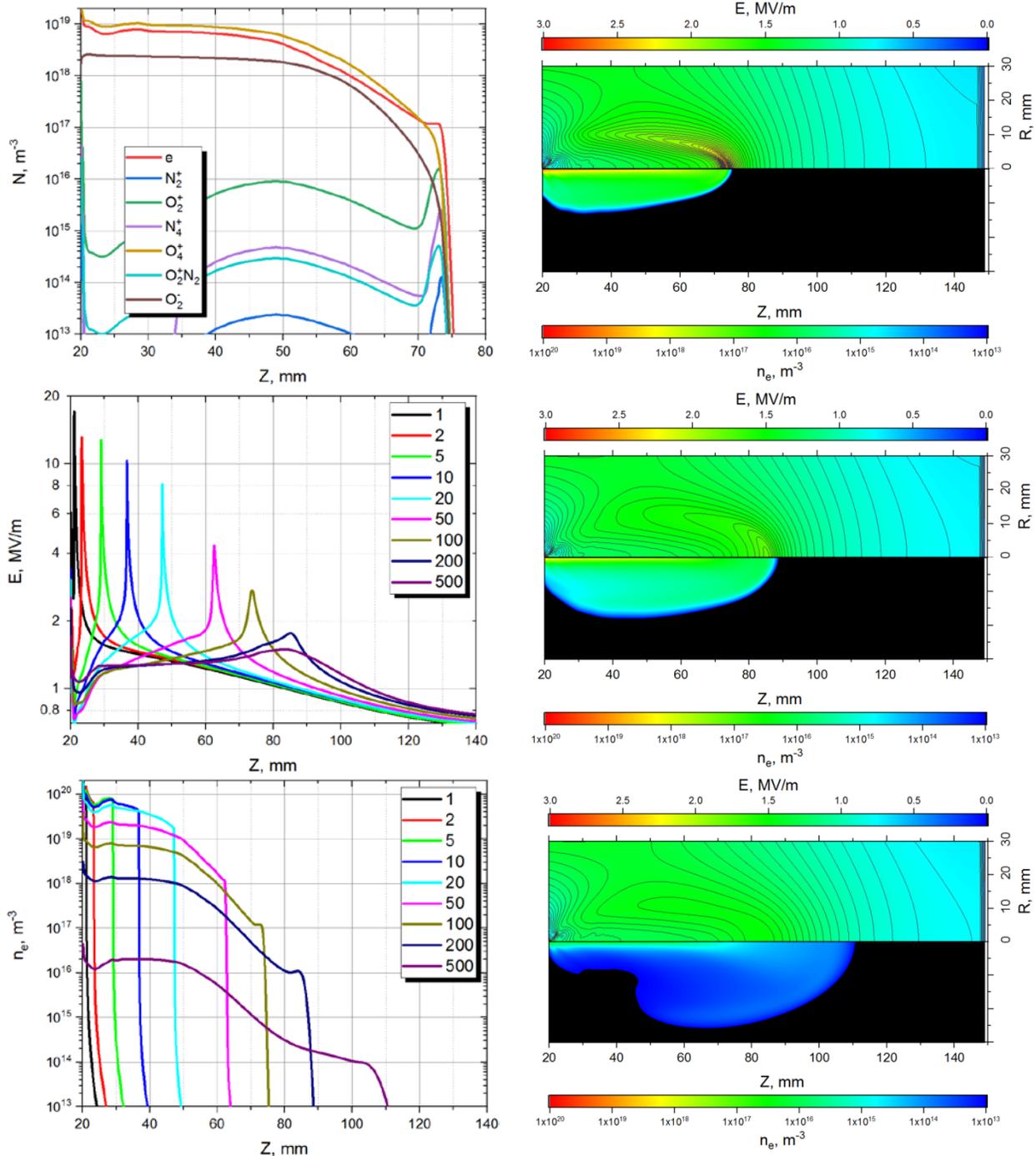

Figure 8. Development of a negative streamer in uniform air. $U_{max} = -150$ kV, $P = 760$ Torr, $T = 300$ K. Left column: Axial profiles of the concentration of the main ions at $t = 100$ ns, and temporal evolution of axial profiles of the electric field and electron concentration. The time scale is shown in nanoseconds. Right column: spatial distributions of electric field and electron concentration at $t = 100$, 200 and 500 ns.

*E/n* ~ 55 Td. The presence of a practically immobile significant space charge reduces the electric field around the high-voltage electrode, preventing from streamer reinitiation.

Thus, regardless of the geometry of the discharge gap, the amplitude of the applied voltage, and the distribution of the gas density in the gap, the positive streamer demonstrates the same behavior in the deceleration phase. With a decrease in the head potential, a positive streamer tends to reduce the radius of curvature of the leading ionization wave in order to maintain the electric field and the rates of ionization and photoionization. However, as the radius of curvature of the streamer head decreases to values comparable to or less than the path length of the photoionizing VUV radiation in the gas, the efficiency of photoionization begins to decrease. This forces the streamer to further reduce the head radius, further increasing the electric field. This process is a classical example of instability development that leads to a singular solution in the fluid approximation.

On the contrary, with a decrease in the head potential, a negative streamer turns out in the a regime when the dominant role in its propagation is played by the electron drift in the head electric field rather than by the gas photoionization ahead of the front. The electron drift allows avalanches moving in the same direction with the streamer to travel to long distances. This ensures a noticeable electron multiplication even in a relatively weak field of a large-radius head.

The "spreading" of the negative streamer leads to a decrease in the peak electric field at the head and in the electron concentration in the channel. Therefore, the streamer stops and forms a wide diffuse space charge. The great difference in the dynamics of deceleration of positive and negative streamers can be analyzed analytically, which will be discussed in the next Section.

## Discussion
### Qualitative theoretical analysis of streamer development in a deceleration mode

The relationship between the ionization and photoionization processes and their influence on the motion of a positive streamer was qualitatively analyzed in [45]. The proposed model considered the 1D electron motion along the streamer axis in the electric field of a semi-spherical streamer head. In this model, the streamer head potential and streamer propagation velocity were related with the head radius. It was shown that electron impact ionization and photoionization controls the streamer propagation under the conditions studied. For positive streamers, the head radius can be determined from the equation that describes the production of VUV photons at the streamer head and the development of an avalanche from a single photoelectron in front of the ionization wave [45]:

$$\frac{1}{2}\frac{p_q}{p+p_q}\int_{r_m}^{\infty}\Psi\big((z_0-r_m)p\big)exp\left(\int_{r_m}^{z_0}\frac{\alpha(z)v_d(z)}{v_s-v_d(z)}dz\right)dz_0 = 1 \qquad (6)$$

where $\alpha(z)$ is the first Townsend coefficient, $v_d(Z)$ is the electron drift velocity, $v_s$ is the streamer velocity, $p$ is the gas pressure, $p_q$ is the quenching pressure for the photo-ionizing states, and $\Psi((z_0-r_m)p)$ is the absorption coefficient of ionizing radiation. This model was generalized to the case of a streamer of any polarity [53] in the limit of "strong" streamers when $v_s \gg v_d$.

For a negative streamer near the stopping point, such an assumption is not valid, since electron avalanches generated by seed electrons ahead of the ionization wave front can move in phase with this wave, producing significant ionization even in a relatively weak electric field. In equation (6), the main limitation is imposed by the denominator of the integrand expression

$$R = \frac{\alpha(z)v_d(z)}{v_s-v_d(z)} \quad , \qquad (7)$$

which tends to infinity when the streamer velocity tends to the electron drift velocity. For positive streamers, the direction of the electron drift is opposite to the streamer propagation and $v_d(z) < 0$ in (7). The electron drift in negative streamers is directed along the streamer development and $v_d(z) > 0$. Thus, for positive polarity, denominator (7) is always positive, while for negative polarity, as the value of $v_s$ approaches $v_d$ during streamer deceleration, the denominator of (7) tends to zero. Physically, as already noted, this means an infinitely long residence time of an electron avalanche in the head electric field. This leads to an unlimited increase in gas ionization in this avalanche and allows the propagation of even weak negative streamers.

To numerically analyze expression (6), it is necessary to have data on the first Townsend coefficient ($\alpha(E/n)$), the electron drift velocity ($v_d(E/n) = \mu_e E$), the quenching pressure for the photo-ionizing states ($p_q$), and the absorption coefficient of ionizing radiation $\Psi((R)\,p)$. The approximations of the data used in this work are shown in Figure 9. The values of the Townsend coefficient and the electron mobility were calculated using the Boltzmann equation in the local approximation [56] and the self-consistent sets of the electron cross sections for $N_2$ and $O_2$ [57]. The averaged absorption coefficient of ionizing VUV radiation, $\Psi((R)\,p)$, was taken based on the approximation of the Penny&Hammert data [58] made in [52]. It was assumed that $p_q = 30$ Torr. A detailed review of the data on photoionization of different gases is given in [59].

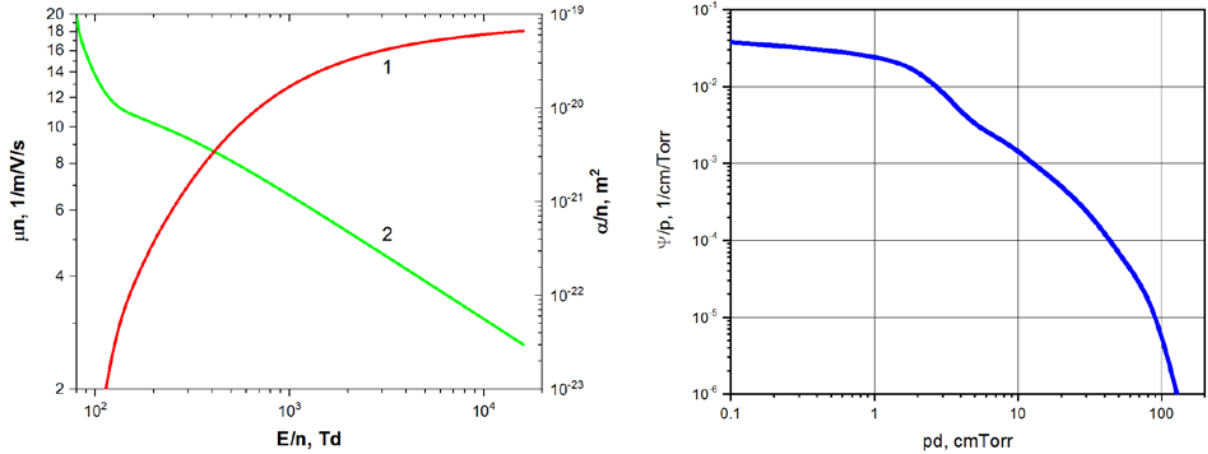

Рис. 9. First Townsend coefficient $\alpha(E/n)$ and electron mobility $\mu_e$ versus $E/N$ and absorption coefficient of ionizing radiation $\Psi(pd)$ versus $(pd)$ for air.

**Field distribution in the front of positive and negative ionization waves**

Using the data in Figure 9, equation (6) allows an estimation of the streamer head radius $r_m$ for various values of the streamer velocity. To estimate the electric field distribution near the streamer head, we utilized the approximation [5]

$$E(z) = E_m \frac{r_m^2}{z^2} \qquad (8)$$

for $z \geq r_m$ and $E(z) = 0$ for $z < r_m$. Here,

$$E_m \approx \frac{U_m}{2r_m} \qquad (9)$$

is the peak electric field strength at the head and $U_m$ is the potential in the region of the maximum field at the front of the ionization wave. Using equations (6), (8), and (9) and taking into account the approximation $U_m \sim U_{head}$, we calculated the streamer velocity for various values of the head potential and a given value of $E_m$. In our case, it was interesting to study the possibility of such a strong decrease in the head radius during streamer deceleration that the electric field at the head exceeds the threshold value for the transition of electrons to the runaway mode. In this case, some electrons near the front of the ionization wave can gain energy corresponding to the potential of the head at this point.

As noted in [41], the uncertainty of the values of such a critical field, calculated under various assumptions, can be quite large. A particularly large uncertainty is observed at low initial electron energies, where, in addition, it is necessary to take into account the electrons energy losses to excitation of the internal degrees of freedom of molecules. For example, in [40], the value of the critical field was taken at the level $E/n$(crit) = 1.7 kTd, while in [41, 60], in the case of a uniform field, a value of $E/n$(crit) ~ 30 kTd was shown for $N_2$. In order to exclude the influence of the choice of a specific value of the threshold field on further conclusions, we performed all subsequent calculations for the value $E/n$(crit) = 3 kTd. The influence of possible uncertainty in the choice of this quantity will be shown below.

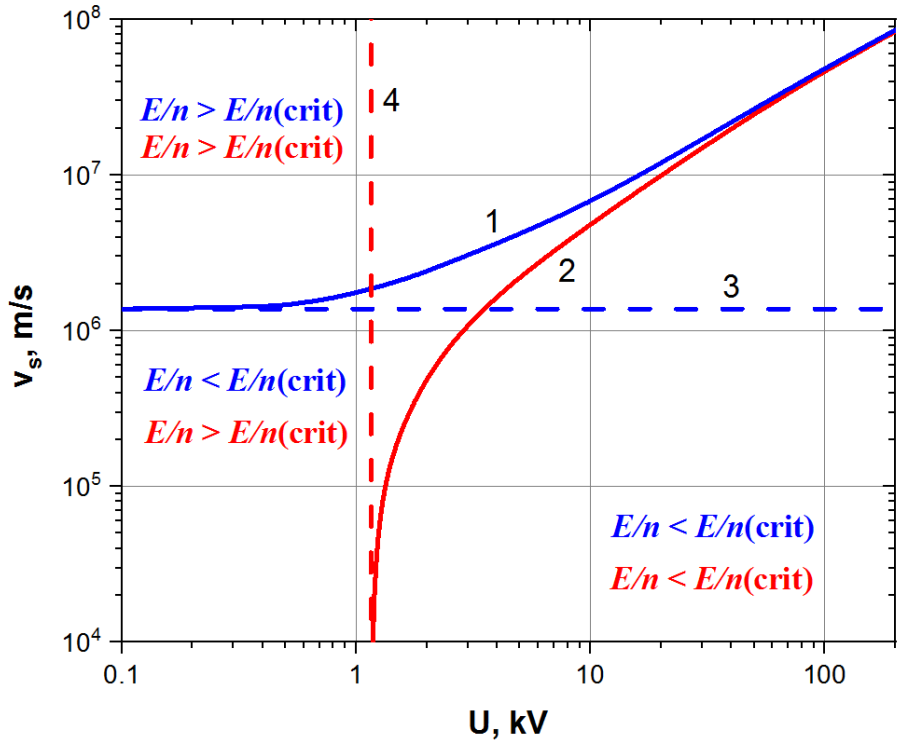

Figure 10. Regions of runaway electron formation at the streamer head in $v_S - U$ space for negative (curve 1) and positive (curve 2) polarities. Curve 3 is the asymptotic velocity of negative streamers at low potential of the head. Curve 4 is the minimum potential of the head for positive streamers, allowing their development without formation of runaway electrons. Isolines $E/n$ = 3 kTd are shown.

Figure 10 shows the results of this calculation when the value $E/n$(crit) = 3 kTd was chosen as the critical value of the reduced electric field at the front of the ionization wave. In this case, a significant amount of the electrons is into the runaway mode and a high-energy electron beam is formed. In Figure 10, the regions in which the reduced electric field at the front of the ionization wave is smaller than the critical one is located to the right of curves 1 and 2. The regions to the left of these curves correspond to the reduced fields exceeding the runaway threshold $E/n$(crit). Thus, for negative streamers, due to the avalanche development in the direction of the streamer propagation, the streamer can develop at any low head potential as long as the ionization rate exceeds the rate of plasma recombination and electron attachment at least in a small region near the front.

A natural limitation on the propagation velocity of negative streamers at very low voltages is the electron drift velocity in the electrical field of the head. When the propagation velocity of the negative streamer is equal to the electron drift velocity, the residence time of the avalanche in the head electric field tends to infinity. This allows air ionization in any electric field when it is above the ionization threshold.

For "strong" streamers developing at high values of the head potential, the contribution of electrons drift to the generation of seed electrons in front of the ionization wave is negligible. Therefore, positive and negative streamers demonstrate the same asymptotic behavior at high $U$ (Figure 10). However, at low head potentials, the difference between negative and positive streamers becomes fundamental. Due to the opposite development of electron avalanches in the positive streamer, it is fundamentally impossible to increase the residence time of the avalanche in the electric field of the streamer head. Therefore, at sufficiently low streamer head potentials, to maintain the balance between photoelectrons and VUV photons produced by the avalanches, which are generated by these photoelectrons, the electric field at the head have to exceed the electron runaway threshold. When the potential of the positive streamer head drops to 1.2 kV, the streamer must either stop completely or reduce the head radius so that the electric field at the head exceeds a threshold value of 3 kTd (Figure 10). Thus, an increase in the local electric field above the electron runaway threshold is the natural final stage of positive streamer deceleration.

**The influence of photoionization intensity on positive streamer deceleration**
Solutions of equation (6) are sensitive to variations in the photoionization intensity in front of the streamer head. In this case, the threshold value of the head potential, which correspond to an increase in the head electric field above the electron runaway threshold, should depend strongly on the intensity of the VUV photon flux from the ionization wave front.

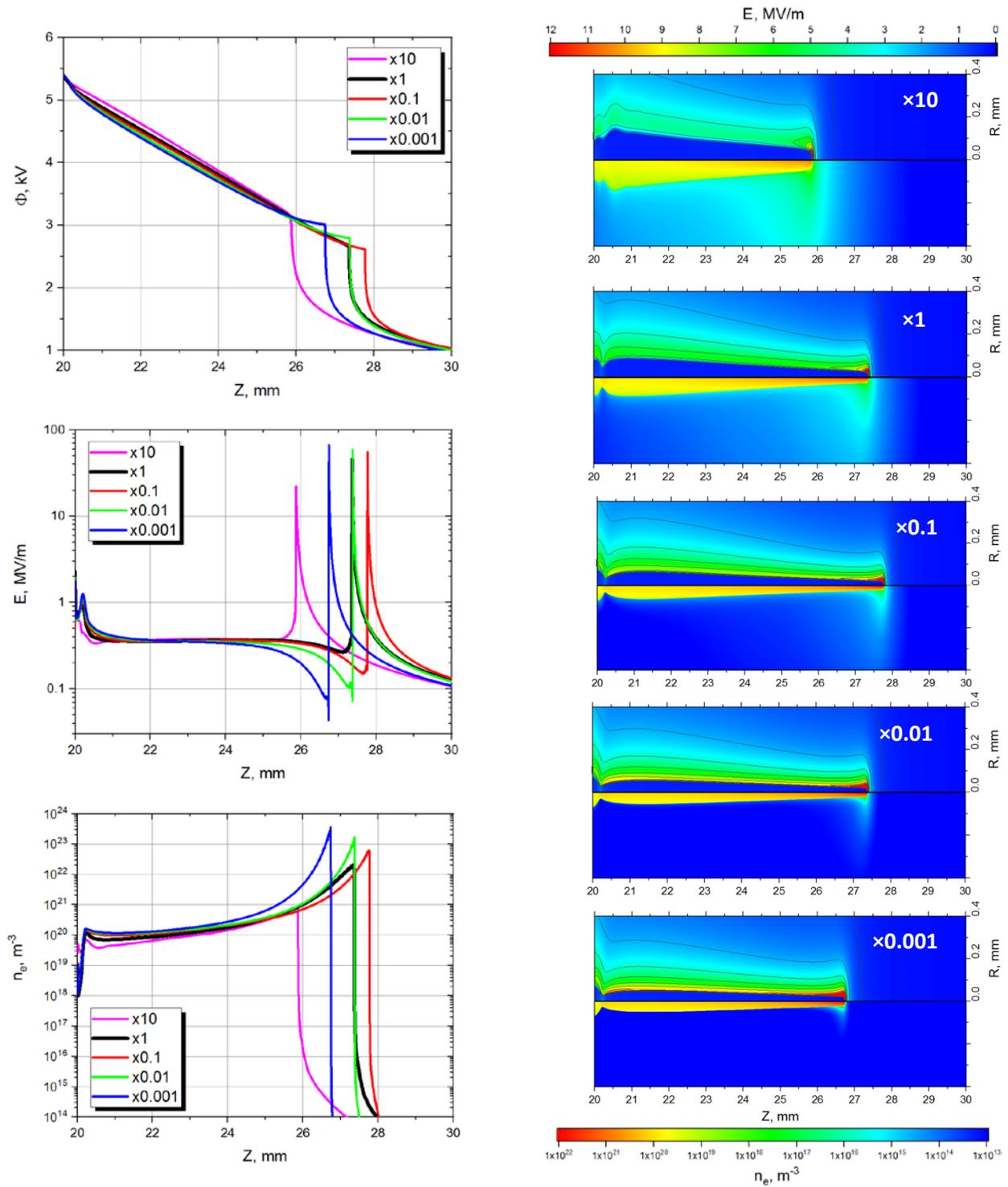

Figure 11. Development of a positive streamer in uniform air for various photoionization rates. $U_{max} = +5.4$ kV, pulse rise time 1 ns. $P = 760$ Torr, $T = 300$ K. Left column: axial profiles of potential, electric field and electron concentration. Right column: spatial distributions of electric field and electron concentration for different photoionization rates. All data are obtained at $t = 30$ ns.

The results of calculations performed with different values of the VUV radiation intensity are shown in Figure 11. The left column shows the axial profiles of the potential, electric field, and electron concentration at $t = 30$ ns. The pulse amplitude was $U = +5.4$ kV and the rise time was 1 ns. The flux of VUV photons was increased by a factor of 10 or decreased by factors 10, 100, and 1000 relative to the standard value of the flux under normal conditions in air. It follows from the calculations that the positive streamer in the deceleration mode responded non-monotonically to the changes in the photoionization intensity. An increase in the flux of VUV photons predictably increased the radius of the streamer channel and slowed down the streamer propagation. A tenfold decrease in the photoionization rate led to a more rapid narrowing of the streamer channel, an increase in the local field at the head, and an acceleration of its propagation as compared with the base case. However, a further decrease in the photoionization rate slowed down the positive streamer, although the electric field at the head and the electron concentration in it continued to increase (Figure 11).

Thus, qualitative analysis on the basis of equation (6) confirms the conclusions made from the 2D numerical simulation about the role of photoionization in positive streamer deceleration and in the temporal evolution of the head electric field.

**The influence of photoionization intensity on generation of runaway electrons**

Figure 12 (left) shows the critical value of the head potential calculated versus the positive streamer velocity. The calculations were made on the basis of equation (6) for different values of the photoionization rate. To the right of the curves, the reduced electric field at the streamer head is lower than the runaway threshold. To the left of the curves, the reduced electric field is greater than the electron runaway threshold in air at atmospheric pressure and runaway electrons are expected to be generated.

Figure 12 (right) shows the critical potential of the positive streamer head at which the electric field at the head reaches the runaway threshold and the generation of high-energy electrons can occur. The calculations were performed for increased and decreased photoionization rates. The critical head potential depends strongly on the rate of air photoionization in front of the streamer head. At high photoionization rates, the streamer does not need to increase the field at the head significantly to continue its propagation. Therefore, an increase in the photoionization rate leads to a decrease in the critical potential of the head from 1160 to 700 V. On the other hand, a decrease in the photoionization rate causes a more intense collapse of the positive streamer and the transition to supercritical electric fields. In this case, the runaway regime occurs at higher head potentials. Thus, an order of magnitude decrease in the photoionization rate increases the critical potential to 1600 V, two orders of magnitude photoionization rate decrease increases the critical potential to 2.1 kV, and three orders of magnitude photoionization rate decrease increases the critical potential to 2.6 kV. Thus, it may be concluded that the presence in the atmosphere of an insignificant impurity with intense absorption bands in the vacuum ultraviolet region of the spectrum can significantly change the deceleration dynamics of positive streamers and the spectrum of high-energy electrons generated at the late stages of streamer deceleration.

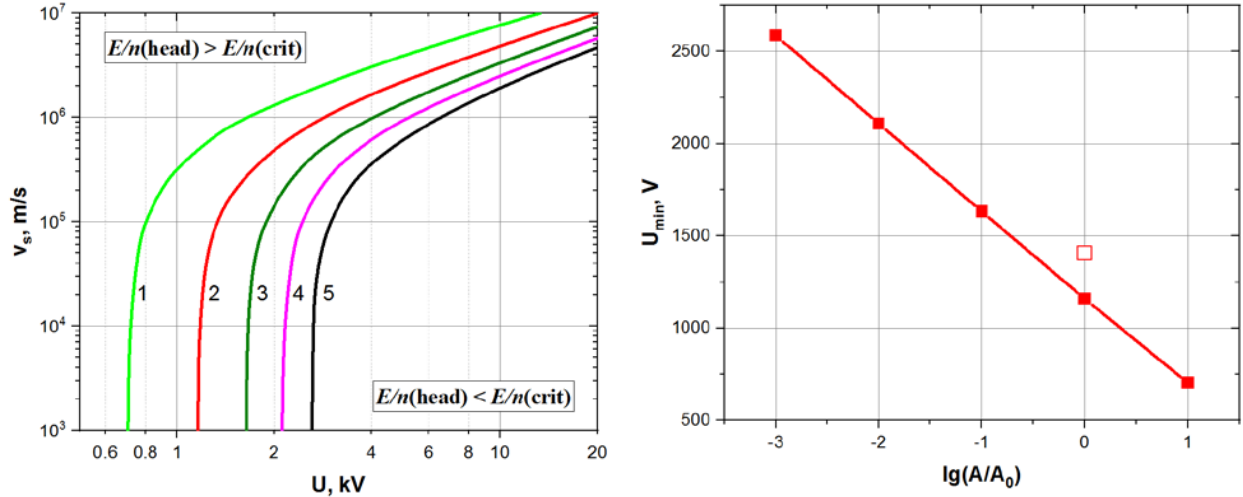

Figure 12. Left: regions of runaway electron formation at the positive streamer head in $v_S - U$ space for different photoionization rates: (1) photoionization rate increased by a factor of 10; (2) basic photoionization rate; and (3)-(5) - photoionization rate is reduced by factors 10, 100, and 1000, respectively. Right: the critical head potential of a decelerating positive streamer, below which runaway electrons are generated, as a function of the relative photoionization rate $A/A_0$ ahead of the ionization wave front. $E/n$(crit) = 3 kTd for all points except open square on the right figure, where $E/n$(crit) = 1.7 kTd was used.

As indicated above, there is a significant uncertainty in the value of the critical field, which causes the transition of electrons to the continuous acceleration regime in the case of low initial electron energies. The effect of this uncertainty is shown in Figure 12, right. The point at the normal value of the photoionization rate was calculated both for $E/n$(crit) = 3 kTd (both throughout this work) and for $E/n$(crit) = 1.7 kTd, which is close to the minimum value of the critical field calculated using only the ionization energy losses and taking into account electrons elastic scattering [40, 41]. It is seen that the choice of the magnitude of the critical field has little effect on the behavior of the decelerating streamer. A slight increase in the potential of the streamer head at the moment of stopping (from 1160 to 1410 V) can lead to a slight increase in the energy of the runaway electrons generated. At the same time, this increase is much less than the decrease in the critical field, which makes it possible to neglect the influence of the uncertainty in the critical value on the conclusions of this work.

**Conclusions**

Simulation of a positive streamer in inhomogeneous air for experimental conditions [15] and comparison of the results obtained with observations have demonstrated the efficiency of the numerical model. Calculations correctly reproduced the streamer deceleration in a gas with varying degree of inhomogeneity. Good agreement was obtained between calculation and experiment for the limiting streamer length, as well as qualitative agreement for the dependence of the effective diameter of the streamer on its length.

It was shown that the positive and negative streamers behave differently during deceleration and stopping. In both cases, deceleration begins with the loss of a significant part of the potential in the lengthening channel. However, the further dynamics of the deceleration process is significantly different. In response to the ionization rate decrease and the decrease of the rate of photoelectron production in front of the head, the negative streamer goes on to the propagation due to the forward drift of electrons. The radius of the ionization wave front increases, whereas the electric field at the streamer head further decreases. As a result, the streamer stops and its head broadens due to the slow drift of free electrons in the residual subcritical electric field. Owing to electron attachment to molecules, the electron-ion plasma is converted to an ion-ion one. Then, the space charge injected into the gap by the streamer remains practically frozen.

A positive streamer cannot respond with forward electron drift when decreasing the head potential and ionization and photoelectron production rates in front of the head. The only advancement mechanism for a positive streamer is to decrease the effective head radius leading to a local increase in the electric field. This allows the streamer to develop, but the decrease in the head radius dramatically reduces the photoionization efficiency. Now the VUV photons run too far from the wave front (on the scale of its radius and the characteristic radius of electric field decrease in front of the head). This forces the streamer to increase the electric field even more, instead of decreasing the electric field at the head as in the case of negative polarity. The radius of the ionization wave is further reduced. As a result of the development of such a positive feedback, the streamer collapses and its radius tends to zero (in a fluid and in the local electric field approximation). In this case, the magnitude of the reduced electric field near the head of the positive decelerating streamer increases strongly. Estimates show that such a locally enhanced electric field becomes larger than the critical field corresponding to the transition of electrons to the runaway mode when the potential of the positive streamer head relative to the surrounding space decreases to ~1.2 kV (in air at atmospheric pressure). Here, a pulsed beam of runaway electrons directed into the channel of the decelerating streamer is generated. Probably, this mechanism accompanying the positive streamer deceleration can explain the recorded bursts of X-ray radiation during the propagation of streamers in long discharge gaps and in the atmosphere. Gas inhomogeneity is not fundamentally important for the runaway electron generation mechanism under consideration. However, it leads to a significant acceleration of the process due to the faster deceleration and stop of the positive streamer as it moves from a less dense medium to a denser one.

## Acknowledgements

This work was supported by DOE grant DE-SC0021330, Texas A&M University/DOE–NETL grant DE-FE0026825 and by DOE grant DE-FE0026825.